\documentclass[twocolumn, pra,showpacs,superscriptaddress]{revtex4}
\usepackage{bbm}
\usepackage{graphicx}
\usepackage{dcolumn}
\usepackage{bm}
\usepackage{CJK}
\usepackage{amsmath}

\begin{document}

\title{Analytical solutions for two heteronuclear atoms in a ring trap}
\author{Xing Chen}
\affiliation{Institute of Physics, Chinese Academy of Sciences,
Beijing 100190, China}
\author{Liming Guan}
\affiliation{Institute of Physics, Chinese Academy of Sciences,
Beijing 100190, China}
\author{Shu Chen}
\email{schen@aphy.iphy.ac.cn}
\affiliation{Institute of Physics,
Chinese Academy of Sciences, Beijing 100190, China}

\begin{abstract}
We consider two heteronuclear atoms interacting with a short-range
$\delta$ potential and confined in a ring trap. By taking the
Bethe-ansatz-type wavefunction and considering the periodic boundary
condition properly, we derive analytical solutions for the
heteronuclear system. The eigen-energies represented in terms of
quasi-momentums can then be determined by solving a set of coupled
equations. We present a number of results, which display different
features from the case of identical atoms. Our result can be reduced
to the well-known Lieb-Liniger solution when two interacting atoms
have the same masses.
\end{abstract}


\maketitle
\date{}


\section{introduction}

Motivated by recent experimental studies on heteronuclear atoms,
mixtures of atomic gas with different masses have recently attracted
lots of attention. Stable mixtures of heteronuclear atoms, for
examples, $^{40}$K and $^6$Li, $^{40}$K and $^{87}$Rb, and $^6$Li
and $^{23}$Na, have already been realized in experiments
\cite{Ospelkaus,Taglieber,Stan,Wille,Inouye,Ferlaino}. The
heteronuclear quantum gas mixture offers a wide range of
possibilities for exploring novel phases and new quantum effects.
Significant efforts have been made on research topics such as
superfluid properties and pairing mechanisms of heteronuclear
fermions \cite{Iskin,Yip,Baranov} and heteronuclear molecule
formation \cite{Ospelkaus,Papp,Deuretzbacher,Voigt}.

The mass ratio between heteronuclear atoms adds a new degree of
freedom for exploring physical effects associated with the mass
imbalance. In general, heteronuclear atomic systems are expected to
behave quite differently than the equal-mass counterpart
\cite{Iskin,Yip,Baranov,Duan,Blume}. To gain deep insight into
properties of heteronuclear systems, some analytical solutions are
especially important. While most of theoretical works on the
many-body heteronuclear systems are within the mean-field theory,
few-body systems can be treated much more precisely within
nonperturbative microscopic frameworks \cite{Blume}. Particularly,
the two-body problem is vital to understand the basic physical
phenomenon from the beginning of quantum mechanics. The two-atom
system composed of identical atoms is exactly solved as one can
separate the center-of -mass and relative motions and thus the
two-particle system turns to a one-body problem
\cite{Lieb,Busch,Shea,Cirone,Arnold}. The problem with different
atoms becomes complicated and generally is hard to get analytical
solutions as it does not separate in center-of-mass and relative
coordinates \cite{Deuretzbacher}. Recently, the scattering of
heteronuclear atom pairs in a one-dimensional optical lattice has
been studied by using the method of Green's function \cite{Molmer}.

In this article, we study the problem of two unequal-mass atoms with
$\delta$-potential interaction in a one-dimensional ring trap with
periodic geometry. The periodic geometry has been experimentally
realized as a circular wave guide or toroidal trap
\cite{ringtrap1,ringtrap2}. For the two-atom system with
$\delta$-potential interaction, it is convenient to treat the
interaction term by separating center-of-mass and relative motion.
However, it is not convenient to apply the periodic boundary
condition by using separated center of mass and relative
coordinates. When the periodic boundary condition is taken into
account, the relative motion is actually coupled to the center-of
-mass motion. In the present work, we analytically solve the problem
of interacting two heteronuclear atoms by taking the
Bethe-ansatz-type wavefunction and considering the periodic boundary
condition properly. While the momentum of center-of-mass motion is a
good quantum number, the relative momentum is coupled to the
center-of-mass momentum due to the periodic boundary condition. Our
results are consistent with the exact Bethe-ansatz solutions of the
Lieb-Linger model when two atoms have the same mass \cite{Lieb}.
With the analytical solution, we obtain the energy spectrum, density
distribution and momentum distribution for different mass ratio. Our
results indicate that they display different features from the
identical two-atom system.

The paper is organized as follows: In Sec.II, we introduce the model
and present a detailed derivation of the analytical solution of the
interacting heteronuclear system. In the limit of equal mass, our
analytical result is compared to the Bethe-ansatz solution of
two-particle Lieb-Liniger model. In Sec.III, we present the energy
spectrum of heteronuclear atoms. In Sec.IV, we show the density
distribution of relative coordinate and the momentum distribution of
heteronuclear atoms. A summary is given in the last section.

\section{The model and its solutions}

We consider an interacting two-atom system in a 1D ring trap
described by the Hamiltonian
\begin{equation}
\hat{H}=-\frac{\hbar ^{2}}{2m_{1}}\frac{\partial ^{2}}{\partial x_{1}^{2}}-%
\frac{\hbar ^{2}}{2m_{2}}\frac{\partial ^{2}}{\partial
x_{2}^{2}}+g\delta (x_{1}-x_{2}),  \label{H1}
\end{equation}%
where $g$ is the interaction strength and the general case with
different masses $m_1$ and $m_2$ is considered. The geometry of the
ring trap enforces the periodic boundary condition
\begin{equation}
\Psi (x_{1}+L,x_{2})=\Psi (x_{1},x_{2}+L)=\Psi (x_{1},x_{2})
\label{PBC}
\end{equation}
to the eigenfunction $\Psi (x_{1},x_{2})$ of Hamiltonian $\hat{H}$.
Introducing relative and center-of-mass coordinates
$x=x_{1}-x_{2}\label{2}$, $
X=(m_{1}x_{1}+m_{2}x_{2})/(m_{1}+m_{2})$, or alternatively $x_{1}=X+\frac{%
m_{2}}{m_{1}+m_{2}}x$, $x_{2}=X-\frac{m_{1}}{m_{1}+m_{2}}x$, we can
rewrite the Hamiltonian Eq.(\ref{H1}) as the separated form
\begin{eqnarray}
\hat{H}&=& \hat{H}_{X}(X)+\hat{H}_{x}(x) \nonumber \\
&=& -\frac{\hbar ^{2}}{2M}\frac{\partial ^{2}}{\partial X^{2}}-\frac{%
\hbar ^{2}}{2\mu }\frac{\partial ^{2}}{\partial x^{2}}+g\delta (x),
\label{H2}
\end{eqnarray}%
where $\mu =m_{1}m_{2}/(m_{1}+m_{2})$ and $M=m_{1}+m_{2}$. The
wavefunctions of the Hamiltonian can take the form of $\Psi
(X,x)=e^{iKX}\varphi (x)$,
where $e^{iKX}$ is the eigenstate of $\hat{H}_{X}(X)$ and the eigenstate of $%
\hat{H}_{x}(x)$ has the form of
\begin{eqnarray}
\varphi (x)&=& (A_{+}e^{ikx}+B_{+}e^{-ikx})\theta (x)
\nonumber \\
& & + (A_{-}e^{ikx}+B_{-}e^{-ikx})\theta (-x), \label{varphi}
\end{eqnarray}%
where $A_{\pm}$ and $B_{\pm}$ are coefficients to be determined. It
is straightforward to get the eigenenergy given by
\begin{equation}
E=\frac{\hbar ^{2}}{2M}K^{2}+\frac{\hbar ^{2}}{2\mu }k^{2}.
\label{E}
\end{equation}
Although formally the center-of-mass and relative parts are
separated, we shall show that the relative quasi-momentum $k$ is
actually coupled with the total momentum $K$ due to the periodic
boundary condition (\ref{PBC}).

The continuous condition of the wavefunction (\ref{varphi}) gives
the following restriction to the coefficients:
\begin{equation}
A_{+}+B_{+}=A_{-}+B_{-}.  \label{w1}
\end{equation}
By integrating the eigen-equation $\hat{H}\Psi (X,x)=E\Psi (X,x)$
from the negative infinitesimal $-\epsilon $ to the positive
infinitesimal $\epsilon $, we get
\begin{equation}
-\frac{\hbar ^{2}}{2\mu }\left( \frac{\partial }{\partial x}\varphi \left(
x\right) _{x=\epsilon }-\frac{\partial }{\partial x}\varphi \left( x\right)
_{x=-\epsilon }\right) +g\varphi \left( 0\right) =0,  \nonumber
\end{equation}%
which leads to
\begin{equation}
ik\left( A_{+}-B_{+}-A_{-}+B_{-}\right) =\frac{\mu }{\hbar ^{2}}g\left(
A_{+}+B_{+}+A_{-}+B_{-}\right) .  \label{w2}
\end{equation}%
The Eq.(\ref{w1}) and Eq.(\ref{w2}) can not uniquely determine all
the parameters. Additional relations can be obtained by considering
the periodic boundary condition (\ref{PBC}). In order to apply the
periodic boundary condition, it is more convenient to work in the
coordinate space of $x_1$ and $x_2$. In terms of $x_{1}$ and
$x_{2}$, the total wavefunction $\Psi (x_{1},x_{2})=\Psi
(X,x)=e^{iKX}\varphi (x)$ can be also represented as
\begin{eqnarray}
& & \Psi (x_{1},x_{2}) \nonumber \\
&=&(A_{+}e^{ik_{1}x_{1}+ik_{2}x_{2}}+B_{+}e^{ik_{1}^{\prime
}x_{1}+ik_{2}^{\prime }x_{2}})\theta (x_{1}-x_{2})+  \nonumber \\
&&(A_{-}e^{ik_{1}x_{1}+ik_{2}x_{2}}+B_{-}e^{ik_{1}^{\prime
}x_{1}+ik_{2}^{\prime }x_{2}})\theta (x_{2}-x_{1}),
\end{eqnarray}%
where $k_{1}=Km_{1}/(m_{1}+m_{2})+k$, $k_{2}=Km_{2}/(m_{1}+m_{2})-k$, $%
k_{1}^{\prime }=Km_{1}/(m_{1}+m_{2})-k$, and $k_{2}^{\prime
}=Km_{2}/(m_{1}+m_{2})+k$. The periodic boundary condition then
enforces
\begin{eqnarray}
A_{+}e^{ik_{1}L} &=&A_{-},  \label{PB1} \\
B_{+}e^{ik_{1}^{\prime }L} &=&B_{-},  \label{PB2} \\
A_{-}e^{ik_{2}L} &=&A_{+},  \label{PB3} \\
B_{-}e^{ik_{2}^{\prime }L} &=&B_{+}.  \label{PB4}
\end{eqnarray}%
By multiplying Eqs. (\ref{PB1}) and (\ref{PB3}) or Eqs (\ref{PB2})and (\ref%
{PB4}), we get $e^{iKL}=1$, which implies
\begin{equation}
KL=2\pi n  \label{BAE1}
\end{equation}
with $n$ the integer. Combining Eqs. (\ref{PB1}) (\ref{PB2}) and (\ref{w1}),
we can represent $B_{+}$, $A_{-}$, $B_{-}$ in terms of $A_{+}$. Substituting
them into Eq. (\ref{w2}), we can derive the following relation
\begin{equation}
k\frac{\hbar ^{2}}{\mu g}=\frac{\sin \left( kL\right) }{\cos \left( K\frac{%
m_{1}L}{m_{1}+m_{2}}\right) -\cos \left( kL\right) }.  \label{BAE2}
\end{equation}
From  Eq.(\ref{BAE2}), it is clear that $k$ is coupled with the $K$
whereas $K$ is uniquely determined by Eq.(\ref{BAE1}). By solving
Eqs. (\ref{BAE1}) and (\ref{BAE2}), we can get solutions of $k $ and
$K$ and therefore eigenvalues of the system by Eq.(\ref{E}).

The above equations can be also represented in terms of the
quasi-momentum $ k_{1} $ and $k_{2}$ via $K=k_{1}+k_{2}$ and $
k=(m_{2}k_{1}-m_{1}k_{2})/(m_{1}+m_{2})$. In terms of $k_1$ and
$k_2$, the eigenenergy is given by
\begin{equation}
E=\frac{\hbar ^{2}}{2m_1}k_1^{2}+\frac{\hbar ^{2}}{2 m_2}k_2^{2}.
\label{E2}
\end{equation}

In order to compare to the case with the same masses, we introduce
$g=\frac{\hbar ^{2}}{m_{1}}c$ and rewrite Eq. (\ref{BAE2}) as
\begin{equation}
\left( 1+\frac{1}{\alpha }\right) \frac{k}{c}=\frac{\sin \left( kL\right) }{%
\cos \left( \frac{KL}{1+\alpha }\right) -\cos \left( kL\right) },
\label{BAE3}
\end{equation}
where $\alpha = m_{2}/m_{1}$ is the mass ratio. When the mass ratio
$\alpha =1$, our results (\ref{BAE1}) and (\ref{BAE2}) should be
consistent with the well-known Bethe-ansatz solution of the
Lieb-Liniger model \cite{Lieb}. Using the above relation
(\ref{BAE3}) and taking $\alpha =1$, we can get
\[
\frac{2k+ic}{2k-ic}=\frac{e^{ikL}-\cos \left( K{L}/{2}\right) }{\cos \left( K%
{L}/{2}\right) -e^{-ikL}}.
\]%
Since $\cos (KL/2)=\cos (\pi n)=\pm 1$, the above equation reduces
to
\[
\frac{k_{1}-k_{2}+ic}{k_{1}-k_{2}-ic}=\pm e^{ikL},
\]
where $+$ (or $-$) corresponds to $n=$even (or odd). By using the relation $%
e^{ikL}=e^{i(k_{1}-k_{2})L/2}=e^{ik_{1}L}e^{-iKL/2}=\pm
e^{ik_{1}L}$, we then have
\begin{equation}
e^{ik_{1}L}=\frac{k_{1}-k_{2}+ic}{k_{1}-k_{2}-ic}, \label{LLBAE}
\end{equation}
which is just the Bethe-ansatz equation of  the two-particle
Lieb-Liniger model \cite{Lieb}. Another Bethe-ansatz equation can be
obtained via $e^{ik_{2}L}=e^{-ik_{1}L}$. When $m_1=m_2$, we have
$k_1^{\prime }=k_2$ and $k_2^{\prime }=k_1$. From Eqs. (\ref{PB1}),
(\ref{PB2}) and (\ref{w1}), we can also get $A_{-}=B_{+}$ and
$B_{-}=A_{+}$. Therefore the wavefunction also has the same form of
the Lieb-Liniger model \cite{Lieb}, which is invariant under the the
exchange of $x_{1}$ and $x_{2}$.
We note that, for the general case with mass imbalance, the
wavefunction is not invariant under the exchange of $x_{1}$ and
$x_{2}$ since $k_{1}^{\prime }\neq $\ $k_{2}$ and $k_{2}^{\prime
}\neq k_{1}$.

\section{The energy spectrum}
Without loss of generality, only the case of $\alpha \geq 1$ shall
be considered. For convenience, in the following calculation we
shall take $\hbar=1$ and $2m_1=1$.  In the units of energy
$\hbar^2/(2m_1)=1$, the energy expression Eq.(\ref{E}) or
Eq.(\ref{E2}) becomes
\begin{equation}
E=(1+\alpha)^{-1} K^{2}+ \left( 1+ {1}/{\alpha }\right) k^2
\label{E3}
\end{equation}
or $ E=k_{1}^{2}+({1}/{\alpha })k_{2}^{2}$. Since the total momentum
$K$ is a conserved quantity determined by Eq.(\ref{BAE1}), the
energy spectrum of the system $E$ can be characterized by different
$K$. The ground state corresponds to $K=0$. Given $K=2\pi n/L$, $k$
can be determined by solving Eq.(\ref{BAE2}) or Eq.(\ref{BAE3}). For
the repulsive case with $c>0$, Eq. (\ref{BAE3}) has only real
solutions which describe the scattering states. However, for the
attractive case with $c<0$, Eq. (\ref{BAE3}) may have imaginary
solutions corresponding to the bound states of the attractive
system.

First we shall consider the case with the total momentum $K=0$. For
$K=0$, Eq.(\ref{BAE3}) becomes
\[
\left( 1+\frac{1}{\alpha }\right) \frac{k}{c}=\frac{\sin \left( kL\right) }{%
1-\cos \left( kL\right) }.
\]%
With similar steps as the equal-mass case, we can get
\begin{equation}
e^{ik L}=\frac{k+ic'}{k-ic'} \label{BAEK0}
\end{equation}
with $c^{\prime }=c/\left( 1+\frac{1}{\alpha }%
\right) $. For $c>0$, $k$ has only real solutions. It is convenient
to solve the real solution of $k$ from the logarithm form of
Eq.(\ref{BAEK0}) given by
\begin{equation}
k L=2\pi ( I-1/2) - 2 \arctan (k/c'), \label{BAEK01}
\end{equation}
where $I$ takes integer. By numerically
solving the $k$ solution, we can get the energy $%
E=\left( 1+\frac{1}{\alpha }\right) k^{2}$ for states with $K=0$.
The ground state energy corresponds to the solution of
Eq.(\ref{BAEK01}) with $I=1$. For $c<0$, Eq.(\ref{BAEK0}) has
imaginary solution $k=i \lambda$. The real solution $\lambda$ can be
obtained by solving
\begin{equation}
e^{- \lambda L}=\frac{\lambda+ c'} {\lambda - c'} , \label{BAEbound}
\end{equation}
which is obtained by inserting $k=i \lambda$ into Eq.(\ref{BAEK0}).
From Eq.(\ref{BAEbound}), one can also check that $\lambda$ has only
solution for $c'<0$. The energy of the bound state can be obtained via $%
E=- \left( 1+{1}/{\alpha }\right) \lambda^{2}$.

To give a concrete example, we consider a two-atom system composed
of $^{40}$K and $^{87}$Rb with the mass ration $\alpha=2.175$. In
Fig.1, we plot the energy spectrum for states with $K=0$. The energy
spectrum for $\alpha=2.175$ is represented by dashed lines. The
energy spectrum of the equal-mass system is also given for
comparison. Here we use the dimensionless energy  $e(\protect\gamma
)=E/(N\protect\rho ^{2})$, where $\rho=N/L$ and $N=2$ is the atom
number. It is shown that the ground energy for the system of $\alpha
>1$ has lower energy than that of the system with $\alpha=1$.
Actually, all the other spectrum corresponding to quantum number $I$
for $\alpha>1$ is below its correspondence for $\alpha=1$.

\begin{figure}[tbp]
\includegraphics[width=3.5in]{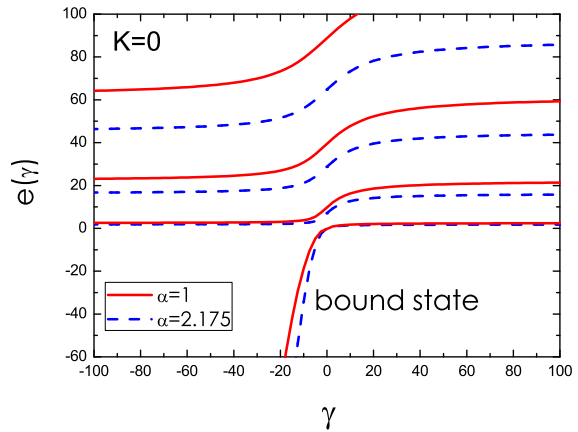}\newline
\caption{(Color online) The dimensionless energy spectrum versus $\protect%
\gamma $ for $K=0$,  $\protect\alpha =2.175$ (dashed line) and
$\protect\alpha =1$ (solid line).  For $\gamma<0$, there exists a
bound state. } \label{fig1}
\end{figure}

\begin{figure}[tbp]
\includegraphics[width=3.5in]{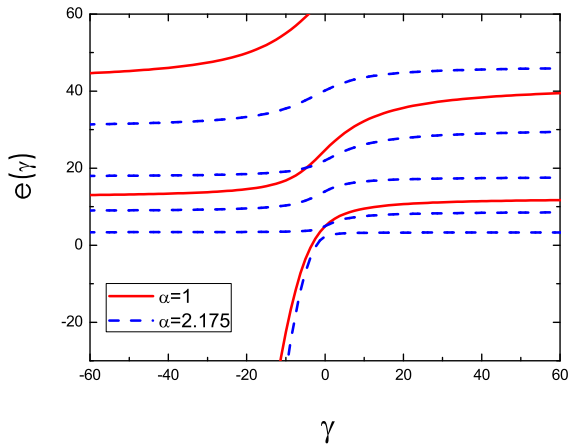}\newline
\caption{(Color online) The dimensionless energy spectrum versus $\protect%
\gamma $ for $K=\pm2\protect\pi/L$,  $\protect\alpha =2.175$ (dashed
line) and $\protect\alpha =1$ (solid line).} \label{fig2}
\end{figure}

Now we consider states with the nonzero total momentum, for example,
$K=\pm \frac{2\pi }{L}$, which correspond to excited states of the
system. When $K \neq 0$, Eq.(\ref{BAE3}) can not be simplified to a
similar form as Eq.(\ref{BAEK0}) or Eq.(\ref{BAEK01}) for an
arbitrary $\alpha$. Nevertheless, formally we have
\begin{equation}
kL=2\pi I -2 \arctan \frac{k}{c^{\prime }}- 2 \arctan [\tan \frac{kL}{2%
}\cot ^{2}\frac{KL}{2(1+\alpha )}], \label{BAEK}
\end{equation}
where $I$ takes integer. When $K=0$, it reduces to
Eq.(\ref{BAEK01}). The relative momentum can be obtained by
numerically solving Eq.(\ref{BAE3}) or (\ref{BAEK}) with $K=\pm
\frac{2\pi }{L}$
and the energy is given by $E=\left( 1+\alpha \right)^{-1} \left(
\frac{2\pi }{L}\right) ^{2}+\left( 1+\frac{1}{\alpha }\right)
k^{2}$. The energies for $K=2 \pi /L$ and $K=-2 \pi /L$ are
degenerate. In Fig.2, we display energy spectrums of states with
$K=\pm 2 \pi /L$ for $\alpha=2.175$ and $\alpha=1$. It is shown that
the energy spectrum for $\alpha>1$ is always below the corresponding
spectrum for $\alpha=1$.

The effect of mass ratio on energy spectrum can be understood from
two aspects. First, we can see that coefficients in Eq.(\ref{E3})
decrease as the mass ration $\alpha$ increases. On the other hand,
while $K$ in Eq.(\ref{E3}) is independent of the mass ratio, $k$ is
related with $\alpha$ via the nonlinear equation (\ref{BAE3}). It is
not easy to analytically analyze the influence of the mass ratio on
$k$ from Eq.(\ref{BAE3}). However, for the case of $K=0$, from
Eq.(\ref{BAEK0}) or Eq.(\ref{BAEK01}), we notice that the mass ratio
$\alpha $ affects the solution of $k$ by effectively changing the
interaction strength via $c^{\prime }$ and thus $k$ increases with
the increase in $\alpha$.


For the attractive case with $c<0$, it is worthy of indicating that
there exists a bound state for each set of states characterized by
$K$ when the interaction exceeds a threshold $c_{t}$. The threshold
of $\gamma_t = c_{t}/\rho$ for the appearance of bound state is
given by
\begin{equation}
\gamma _{t} = \frac{1}{2}\left( 1+\frac{1}{\alpha }\right) \left[
\cos \left( \frac{KL}{1+\alpha }\right) -1\right].
\end{equation}
For $K= 0$, we have $\gamma _{t}=0$. This implies that the atoms
form bound state for arbitrary weak attraction $\gamma <0$. For $ K
= 2\pi /L$, we have $\gamma _{t} < 0$. In this case, atoms can form
bound state only for $\gamma < \gamma _{t} $.

\section{density and momentum distribution}

To show the influence of the interaction on the wavefunction, we
plot the ground state density distribution for the relative motion
defined as $ \rho (x)=\varphi ^{*}(x)\varphi (x) $ in Fig.4. The
density distribution displays a cusp at $x=0$ due to the repulsive
interaction when two atoms coincide. The relative motion density
distribution at $x=0$ decreases as the interaction strength
increases. The density distribution at $x=0$ for the case of
$\alpha>1$ is below that of the equal-mass case as a consequence of
increasing $c'$ effectively. In the limit of $c \rightarrow \infty$,
the wave function $\Psi(x_1,x_2)$ vanishes whenever $x_1=x_2$.

\begin{figure}[tbp]
\includegraphics[width=3.5in]{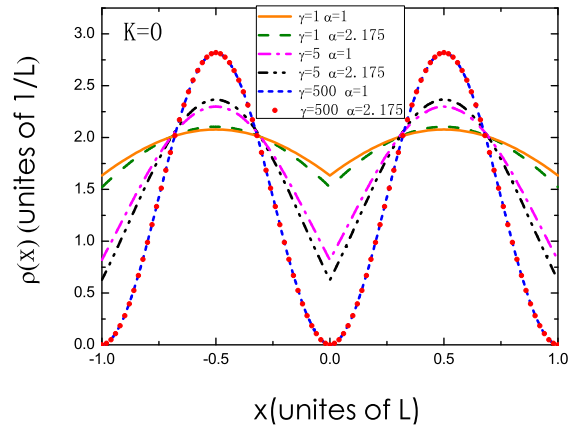}\newline
\caption{(Color online) The probability density distribution of
relative motion for the ground state with $\protect\gamma=1$ and
$\protect\alpha=1$ (dash
dot line), $\protect\gamma=1$ and $\protect\alpha=2.175$ (dashed line), $\protect%
\gamma=5$ and $\protect\alpha=1$ (short dashed line), $\protect\gamma=5$ and $%
\protect\alpha=2.175$ (dashed dotted dot line), $\protect\gamma=500$
and $\protect\alpha =1$ (short dashed dotted line),
$\protect\gamma=500$ and $\protect\alpha=2.175$ (red dotted). }
\label{fig3}
\end{figure}

The momentum distribution for atom with mass $m_{1}$ can be
represented as
\[
n_{1}(p_1)=\int \Psi ^{* }(p_{1},p_{2})\Psi (p_{1},p_{2})dp_{2},
\]%
where $\Psi (p_{1},p_{2})$ is the Fourier transformation of $\Psi
(x_{1},x_{2})$. The definition of momentum distribution $n_{2}(p)$
for the atom with mass $m_2$ is similar. For the equal-mass case,
the momentum distribution for atom $1$ or $2$ should be the same
since the wave function $\Psi(x_1,x_2)$ is invariant under the
exchange of $x_1$ and $x_2$. However, the unequal-mass atomic system
will show different features since the wave functions are not
necessary to be symmetrical for exchanging two unequal mass atoms
except of the case of $K=0$. For $K=0$, we find that the
wavefunction is invariant under exchange of $x_1$ and $x_2$ even for
the unequal-mass case, and thus we have $n_{1}(p)=n_2(p)$. In Fig.4,
we display the ground state momentum distribution for the
unequal-mass system with $\alpha =2.175$ and the equal-mass system.
An obvious peak around the zero momentum position can be observed.
With the increase in the interaction strength, the momentum
distribution tends to spread out widely and the height of the peak
decreases. There is no obvious difference for momentum distributions
of systems with $\alpha=2.175$ and $\alpha=1$ as the influence of
$\alpha $ can be attributed to the effective interaction $c'$.

\begin{figure}[tbp]
\includegraphics[width=3.5in]{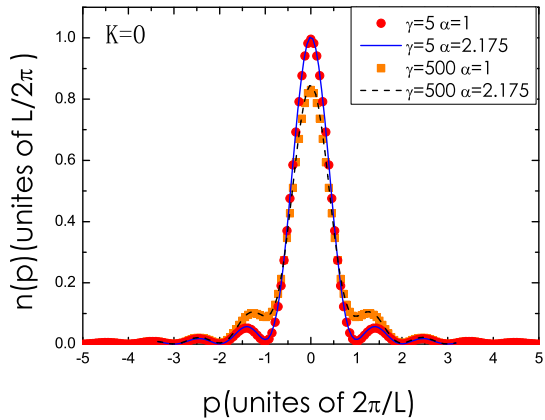}\newline
\caption{(Color online) Single particle momentum distribution of the ground state
for $\protect\gamma =5$ and $\protect\alpha =1$ (dot), $\protect%
\gamma =5$ and $\protect\alpha =2.175$ (solid line), $\protect\gamma =500$ and $%
\protect\alpha =1$ (square), $\protect\gamma =500$ and $\protect\alpha =2.175$%
 (dashed line). } \label{fig4}
\end{figure}

\begin{figure}[tbp]
\includegraphics[width=3.5in]{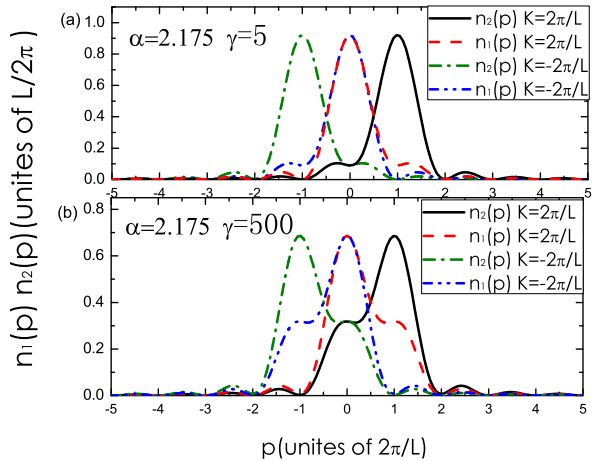}\newline
\caption{(Color online) Single particle momentum distribution of the
lowest state with $K=\pm 2\pi/L$ and $\alpha=2.175$ for (a) $
\protect\gamma =5$ and (b) $\protect\gamma =500$.} \label{fig5}
\end{figure}

Next we discuss the momentum distribution for the lowest state with
$K=\pm 2\pi /L$. For these cases, unequal mass will have significant
impact on the wavefunction and momentum distribution as the wave
functions are not invariant under the exchange of two heteronuclear
atoms. Consequently, the momentum distributions $n_1(p)$ and
$n_2(p)$ are different when $m_1 \neq m_2$. In Fig.5, we display the
momentum distributions for the lowest states of the systems with
$K=\pm 2\pi /L$ and $\alpha=2.175$. For $K=2\pi /L$, the momentum
shows the peak of $n_{1}(p)$ is around $0$ and that of $n_{2}(p)$ is
around $2\pi /L$. The momentum distribution for the noninteracting
system with the same $K$ and $\alpha$ has a similar structure as the
lowest state with $K=2\pi /L$ is composed of the product of single
atom states characterized by $k_{1}=0$ and $k_{2}={2\pi }/{L}$. For
$K=-2\pi /L$, the momentum distributions show that the peak of
$n_{1}(p)$ is around $0$ and that of $n_{2}(p)$ is around $-2\pi
/L$. With the increase in the strength of repulsive interactions,
the momentum distributions become broader and the height of the main
peak shrinks. As a comparison, in Fig.6 we display the momentum
distribution for the lowest states of the equal-mass systems with
$K=\pm 2\pi /L$. Due to the exchange symmetry of the wave function,
the momentum distributions $n_1(p)=n_2(p)$ when $m_1=m_2$.  For
$K=2\pi /L$, the momentum distribution displays two peaks around $0$
and $2\pi /L$, while the distribution has two peaks around $0$ and
$- {2\pi }/{L}$ for $K=-2\pi /L$. With the increase in the strength
of the repulsive interaction, the height of the peaks also shrinks
and the momentum distributions become broader.

\begin{figure}[tbp]
\includegraphics[width=3.5in]{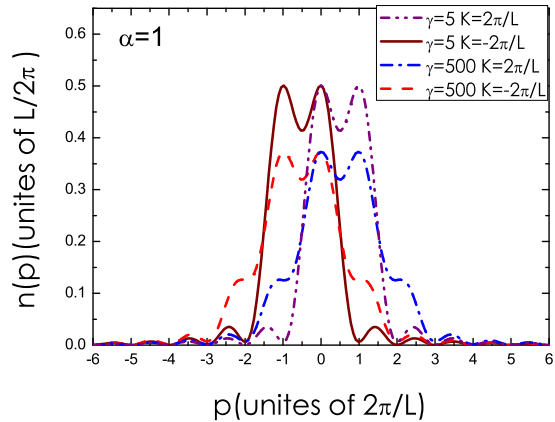}\newline
\caption{(Color online) Single particle momentum distribution of the
lowest state with $K=\pm 2\pi/L$ and $\alpha=1$ for $ \protect\gamma
=5$ and $\protect\gamma =500$.} \label{fig6}
\end{figure}

\section{conclusions}

In summary, we have exactly solved the problem of two heteronuclear
atoms interacting with a short-range $\delta$ potential in a ring
trap. For the general case with an arbitrary mass ration $\alpha$,
the relative momentum is coupled with center-of-mass momentum due to
the periodic boundary condition. By taking the generalized
Bethe-ansatz-type wave function, we have presented a detailed
derivation of the analytical solution. While $\alpha =1$, our
solution reduces to the Bethe-ansatz solution of the two-particle
Lieb-Liniger model. We have applied our analytical results to
studying the energy spectrum, density distribution and momentum
distributions of the heteronuclear system, which differ from the
case of identical atoms.

\textbf{Acknowledgments.--} This work is supported by the NSF of
China under Grants No.10821403 and No.10974234, and National Program
for Basic Research of MOST.

\end{document}